# Transition guide d'onde rectangulaire vide–guide chargé pour réaliser un dispositif de mesure à base de jet électromagnétique

A. DEUBAIBE[1,2], S. NDOUWE[1], B. MAHAMOUT MAHAMAT[1], D. DARI YAYA[1] and B. SAUVIAC[3]

[1] Institut National Supérieur des Sciences et Techniques d'Abéché, Tchad
[2] Université de Ndjaména, Tchad
[3] Université Jean Monnet Saint-Etienne, CNRS, Institut d'Optique Graduate School, Laboratoire Hubert Curien UMR 5516, F-42023, SAINT-ETIENNE, France
antonydeb1982@gmail.com

*Résumé* — Des sphères, des cylindres, des cônes, des cuboïdes ont été utilisés dans la littérature pour générer des jets photoniques à l'arrière de ces structures lorsqu'elles sont éclairées par des ondes planes. Nous travaillons à la généralisation (que nous appelons jet électromagnétique) de ce concept en structure guidée dans le domaine microonde. Le jet électromagnétique peut être obtenu en sortie d'un guide d'onde rectangulaire chargé de diélectrique. Pour pouvoir utiliser le jet électromagnétique comme outil de mesure localisé, il est donc nécessaire de s'intéresser à l'adaptation entre un guide d'onde rectangulaire vide et un guide chargé. C'est l'objet du travail présenté ici, où deux solutions sont comparées.

## I. INTRODUCTION

En 2004 Chen et al [1] montrent la possibilité de générer une focalisation en champ proche faiblement divergente et caractérisée par une largeur à mi-hauteur inférieure à la longueur d'onde. A la suite de ces travaux, des sphères, des cylindres, des cônes, des cuboïdes ont été utilisés dans la littérature pour générer des jets photoniques à l'arrière de ces « particules » lorsqu'elles sont éclairées par des ondes planes. En dépassant la limite de diffraction, ce phénomène, initialement nommée jet photonique, offre un très grand nombre d'opportunité.

En généralisant le principe à d'autres gammes de fréquences [2], sous la dénomination de « jet électromagnétique », et générant le phénomène à l'aide de structures guidées [3] il est possible d'imaginer des dispositifs de mesure de type contrôlent non destructif. C'est dans cette optique que nous envisageons mettre en évidence le phénomène à l'aide de guides d'onde rectangulaires.

La structure à l'étude est un guide d'onde rectangulaire multimodes rempli de téflon et excité avec le mode d'ordre supérieur TE01 à 5GHz. Pour réaliser le dispositif expérimental complet, il est nécessaire d'exciter la structure avec un guide d'onde vide classique. Il est donc impératif de travailler l'adaptation entre le guide d'onde rectangulaire vide et le guide d'onde rectangulaire à téflon. Dans la littérature, l'adaptation entre guides d'onde électromagnétique de différentes natures a généré de très nombreuses études depuis de très nombreuses années. Ce thème reste toujours d'actualité [4], dès qu'il s'agit de coupler deux structures de natures différentes entre guides d'onde de tailles différentes [5], de formes différentes [6] ou permittivités différentes [7]. Dans notre cas, il s'agit de raccorder deux guides d'ondes rectangulaires WR430 chargés de permittivités différentes. Nous présenterons dans cet article, deux façons de réaliser l'adaptation.

## II. JET ÉLECTROMAGNÉTIQUE

La figure1 montre la structure permettant de générer le jet électromagnétique. Selon la forme de l'embout, il est possible de jouer sur les caractéristiques du jet électromagnétique généré dans l'air.

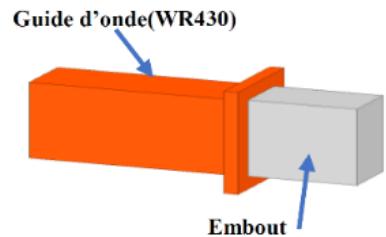

Figure 1: Guide d'onde chargé de téflon

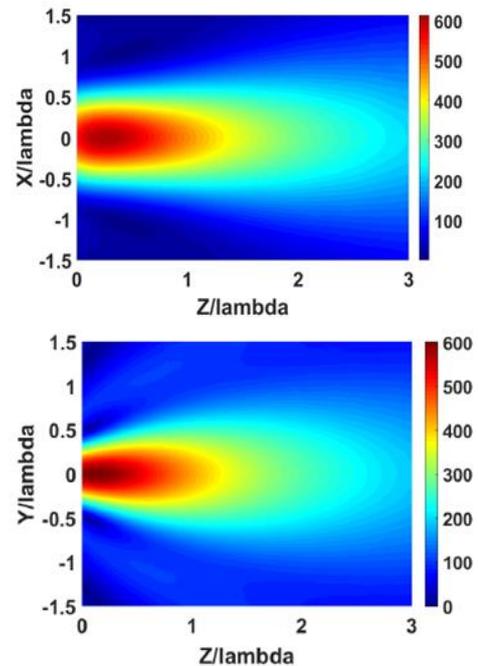

Figure 2: Focalisation du champ électrique en sortie de la structure en fonction de l'éloignement de l'embout (suivant Z) et selon le petit côté du guide (axe Y) et le grand coté (axe X) du guide d'onde rectangulaire [8].

Lorsque le guide est excité par le bon mode avec le bon embout et à la bonne fréquence, il est possible de focaliser l'énergie électromagnétique à proximité de la sortie en zone de champ proche (figure 2). On observe alors un phénomène de propagation stable localement sur quelques longueurs d'onde.

Afin de réaliser un dispositif expérimental facilement exploitable, nous souhaitons pouvoir utiliser des éléments de guide d'onde complètement standards (transition guide coaxial-guide rectangulaire vide, tronçon de guide d'onde rectangulaire vide, tronçon de guide d'onde rectangulaire que l'on pourra remplir de diélectrique dont une partie émergera de la structure – Fig. 1). Seul l'embout, sa forme, ses dimensions devront permettre de maîtriser et générer le jet électromagnétique. Nous allons donc dans la suite travailler au couplage de ce dispositif avec un guide d'onde rectangulaire vide.

### III. GUIDE D'ONDE WR430 ET SIMULATIONS

Le guide d'onde rectangulaire utilisé pour notre étude est du type WR430. Comme il s'agit d'un guide standard, ses dimensions sont a= 54.61mm, b=109.22mm. La fréquence d'étude est fixée à 5GHz. On notera qu'à cette fréquence, la propagation dans le guide est potentiellement multimodale : 10 modes possibles dans le guide vide et 21 dans un guide chargé de Téflon ayant les mêmes dimensions. Les trois premiers modes (TE10, TE20 et TE01) ont respectivement des fréquences de coupure de 1.39 GHz et 2.738 GHz dans le guide vide et 0.916GHz et 1.872GHz dans le guide chargé de Téflon, les modes TE20 et TE01 ayant des fréquences de coupure dégénérées.

Le problème électromagnétique sera traité à l'aide de la méthode des éléments finis. Dans les simulations, le Téflon utilisé dans le guide chargé, sera caractérisé par une permittivité relative de 2,1.

### IV. TRANSITION DANS LE GUIDE D'ONDE RECTANGULAIRE WR430

Pour traiter le problème de l'adaptation entre les deux guides, nous proposons deux types de transition à intercaler entre le guide vide et le guide chargé de Téflon. Dans tous les cas, il s'agit de produire un changement progressif d'indice entre les deux guides d'ondes rectangulaires.

La première version, est une adaptation en pyramide (figure 3). Elle consiste à implanter une pyramide de longueur 100mm à base rectangulaire en Téflon, entre le guide d'onde vide et le guide chargé.

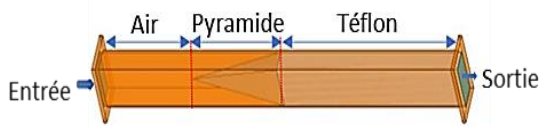

Figure 3: transition en pyramide entre deux guides d'ondes rectangulaires WR430(vide-téflon) : schéma 3D.

La deuxième version est présentée sur la figure 4. On introduit une encoche en téflon (en forme de pince à linge) entre les deux guides. Le diélectrique remplit de façon homogène le petit côté du guide, alors que sur le grand côté, on observe un remplissage progressif de Téflon. La pointe de l'encoche se situe au centre du grand côté du guide avec une longueur de 120mm

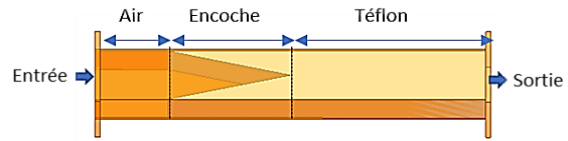

Figure 4 : transition en encoche entre deux guides d'ondes rectangulaire WR430(vide-téflon) : schéma 3D.

### V. RÉSULTATS ET DISCUSSIONS

#### A. Transition pyramidale

La structure de la figure 3 a été optimisée pour obtenir des pertes d'adaptation minimales. La pyramide présente une longueur de 100mm. Les figures 5a, 5b et 5c donnent les résultats des paramètres S des trois premiers modes.

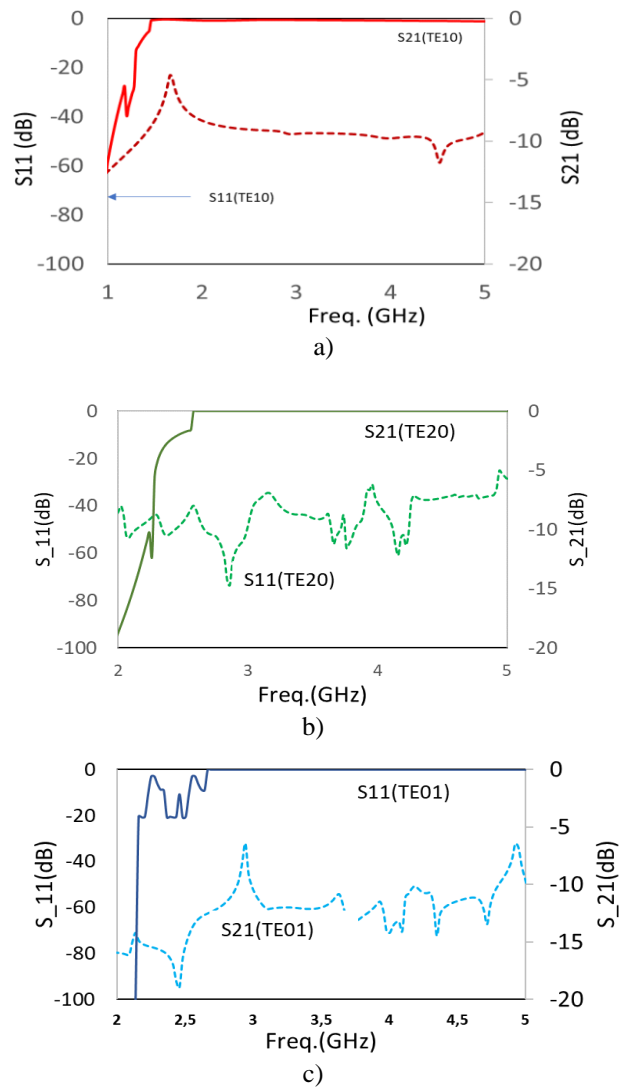

Figure 5 : Réponse de la transition pyramidale pour le mode TE10 (a), TE20 (b) et TE01(c).

Nous observons d'abord sur ces trois figures, que les ondes commencent à se propager respectivement aux fréquences 1,39GHz, 2,78GHz et 2,78GHz, qui correspondent aux fréquences de coupure des modes TE10, TE20 et TE01 respectivement.

Les trois figures montrent que l'onde se propage avec très peu de pertes et une réflexion inférieure à -20dB à 5GHz pour chaque cas. On remarque également que cette réflexion est extrêmement faible pour le mode TE01 (<-50dB).

*B. Transition en forme d'encoche*

De manière similaire, l'optimisation des dimensions a été réalisée pour la transition en forme d'encoche. La longueur d'encoche retenue est de 150mm. Les figures 6a, 6b et 6c donnent les paramètres S de la structure, excitée respectivement par les modes TE10, TE20 et TE01.

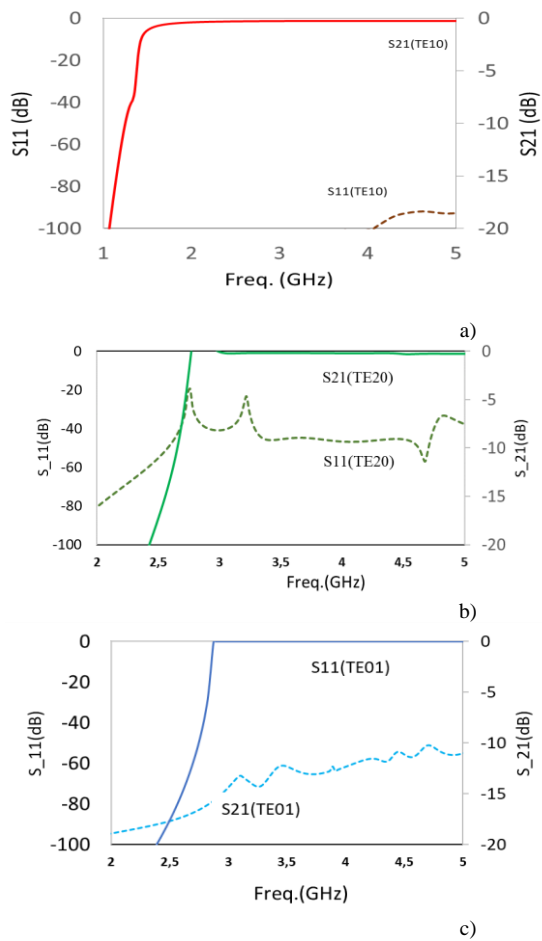

Figure 6 : Paramètres S de la structure avec transition en forme d'encoche pour les modes TE10 (a), TE20 (b) et TE01(c).

Comme le cas précédent nous pouvons observer pour les basses fréquences, l'effet des fréquences de coupure. Pour cette structure, l'adaptation semble meilleure. Les 3 modes présentent des niveaux de réflexion inférieurs à -50dB à 5GHz. C'est le mode TE10 qui présente les meilleurs niveaux de réflexion avec une réflexion inférieure à -90dB sur toute la bande.

On note que la transition en forme d'encoche donne une adaptation légèrement meilleure pour le mode fondamental, alors que la transition en forme de pyramide donne une adaptation sensiblement meilleure pour le 3$^{ème}$ mode. En conclusion d'une manière générale, nous observons qu'en optimisant correctement la longueur de la zone d'adaptation l'onde se propage avec un minimum de pertes, avec une réflexion inferieure à -40dB à 5GHz et des pertes d'insertion inférieures à 0,25dB. Les 2 approches peuvent donc faire l'affaire mais il semble que la structure la plus facile à réaliser technologiquement soit la pyramide.

## VI. REPARTITION DU CHAMP ELECTRIQUE APRES LA TRANSITION

Afin de vérifier que les transitions conservent bien les modes en sortie, nous présentons la cartographie des champs électriques des modes TE10, TE20 et TE01 à la sortie du guide chargé de Téflon. Les résultats sont donnés sur la figure 7 pour la transition pyramidale.

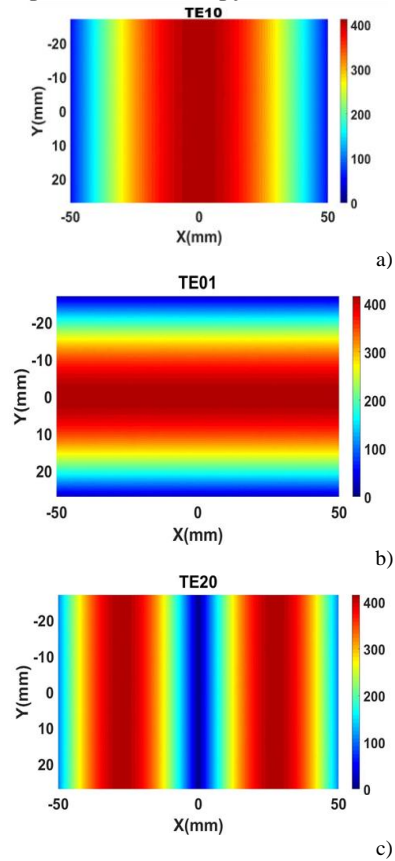

Figure 7 : cartographie du champ électrique des modes (a) TE10 (b) TE01 et (c) TE20 dans le guide d'onde rectangulaire chargé. Cas de la pyramide.

Comme on peut le noter sur la figure 7, la répartition du champ électrique dans la section transverse du guide chargé est parfaitement respectée, ce qui valide complètement la méthode d'adaptation.

## VII. REPONSE DE LA STRUCTURE COMPLETE

Nous avons maintenant simulé l'ensemble de la structure et observé le jet généré en sortie de la structure (figure 8). Nous avons comparé les jets électromagnétiques générés en sortie par le mode TE01 dans le cas :
- D'un guide chargé de Téflon depuis la source (8a)
- D'une jonction guide vide-guide chargé sans transition (8b)
- D'une jonction avec adaptation pyramidale (8c)
- D'une adaptation en forme d'encoche (8d)

On note que dans les différentes configurations, l'embout est bien dimensionné puisqu'il permet la génération d'un jet bien formé à l'extérieur de la structure. Les différents jets obtenus ont sensiblement la même allure puisqu'ils sont générés par un mode ayant la même répartition spatiale à l'intérieur du guide. On peut toutefois remarquer que la longueur suivant z du jet est réduite par rapport au cas idéal (8a). Ensuite, le point le plus notable est la différence d'amplitude du champ dans le jet. Dans le cas du guide chargé unique (8a) le niveau de champ est de l'ordre de 800 V/m au foyer du jet. S'il n'y a pas de système d'adaptation (6b), le niveau est divisé par 2. Enfin dans le cas de nos 2 transitions (8c) et (8d) le niveau observé est de l'ordre de 770 V/m.

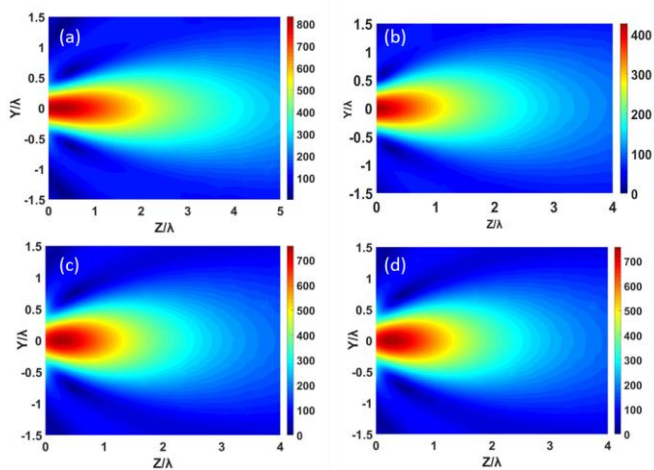

Fig. 8. Cartographie du champ électrique du jet électromagnétique en sortie de l'embout (mode TE01, 5GHz). (a) guide chargé de Téflon depuis la source (b) guide chargé de Téflon excité par un guide vide sans transition (c) guide chargé de Téflon excité par un guide vide avec transition pyramidale (c) guide chargé de Téflon excité par un guide vide avec transition de type encoche.

## VIII. CHOIX DE MODE DE PROPAGATION

Afin de générer le jet électromagnétique avec une concentration homogène sur axe x ou y, nous orientons naturellement notre choix vers le mode TE10 ou TE01 car la répartition du champ dans le guide d'onde exerce nécessairement une influence sur la forme du jet électromagnétique en sortie. On note par exemple que l'excitation de l'embout actuellement dimensionné, par le mode TE20 génère un double jet comme cela est présenté à la figure 9. Toutefois, on remarquera que l'intensité au foyer du jet est nettement moindre que dans le cas de la figure 8.

Une sélection judicieuse des modes d'excitation peut donner ainsi une grande latitude quant à la forme et la répartition du jet en sortie.

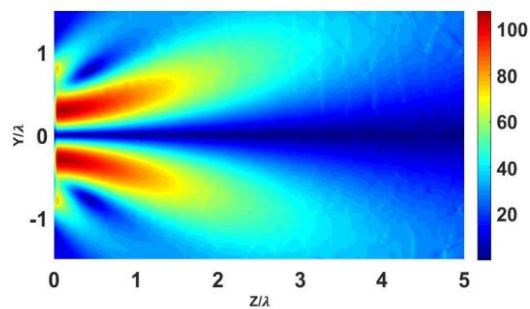

Fig. 9. Cartographie du champ électrique du jet électromagnétique en sortie de l'embout (mode TE20, 5GHz).

## IX. CONCLUSION

Notre objectif était de proposer une organisation simple à base de guides d'onde rectangulaires standards pour générer en sortie des jets électromagnétique. Nous avons montré qu'il est possible de proposer des formes simples permettant de réaliser une bonne adaptation avec changement progressif du milieu d'indice entre les deux guides. Il devient alors possible de générer le jet électromagnétique à la sortie de cette structure. La modification de la géométrie de l'embout peut permettre de modifier la forme du jet en sortie et nous avons également montré qu'il était également possible d'agir en sélectionnant les modes d'excitation de la structure. Il faut donc maintenant designer un sélecteur de mode efficace.